# Contact mechanics of fractal surfaces by spline assisted discretization

*Dorian A.H. Hanaor*[*], *Yixiang Gan, Itai Einav*

School of Civil Engineering, University of Sydney, Sydney, NSW 2006, Australia

**Abstract:**

We present a newly developed approach for the calculation of interfacial stiffness and contact area evolution between two rough bodies exhibiting self-affine surface structures. Using spline assisted discretization to define localized contact normals and surface curvatures we interpret the mechanics of simulated non-adhesive elastic surface-profiles subjected to normal loading by examining discrete contact points as projected Hertzian spheres. The analysis of rough-to-rough contact mechanics for surface profiles exhibiting fractal structures, with fractal dimensions in the regime 1-2, reveals the significant effect of surface fractality on contact mechanics and compliance with surfaces of higher fractality showing lower contact stiffness in conditions of initial contact for a given load. The predicted linear development of true contact area with load was found to be consistent with diverse existing numerical and experimental studies. Results from this model demonstrate the applicability of the developed method for the meaningful contact analysis of hierarchical structures with implications for modelling tribological interactions between pairs of rough surfaces.

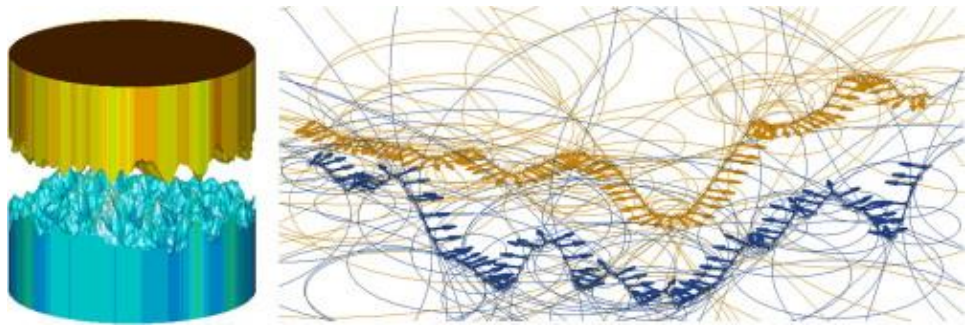

## 1. Introduction:

Interfacial roughness plays a key role in the mechanical behaviour of contacting bodies with significant implications for processes of force transfer, friction, wear and adhesion. The multi-scale structures exhibited by material surfaces govern the extent and distribution of areas of real contact and force-displacement relationships under given conditions of loading. Consequently, the interplay of surface-structure and contact-mechanics has been the subject of extensive research efforts with pioneering studies by Bowden and Tabor (Bowden and Tabor, 1950) and Greenwood and Williamson (Greenwood and Williamson, 1966) spawning a raft of further studies over the past decades (Akarapu et al., 2011; Batrouni et al., 2002; Bhushan, 1998; Paggi and Zavarise, 2011; Patra et al., 2008; Tao et al., 2001). Contact mechanics of rough surfaces are of particular interest for the interpretation of tribological interactions across multiple scales (Buzio et al., 2003b; Carpinteri and Paggi, 2005; Grzemba et al., 2014; Misra and Huang, 2012) with valuable engineering applications ranging from nano-electromechanical systems to tectonic dynamics. Additionally, the structure dependence of contact stiffness and frictional interactions constitutes valuable input in the construction of meaningful multi-body models as used for the analysis of granular matter, geomaterials and powders by discrete element method (DEM) models (Alonso-Marroquín et al., 2013; Luding, 2008).

At the interface between two approaching bodies the transition from non-contact conditions to bulk-type behaviour (elastic or otherwise) occurs through the gradual deformation of asperities (Barber and Ciavarella, 2000; Wriggers and Zavarise, 2002). Consequently the stress-strain behaviour at the interface of contacting elastic bodies generally exhibits a form similar to that shown in Figure 1.

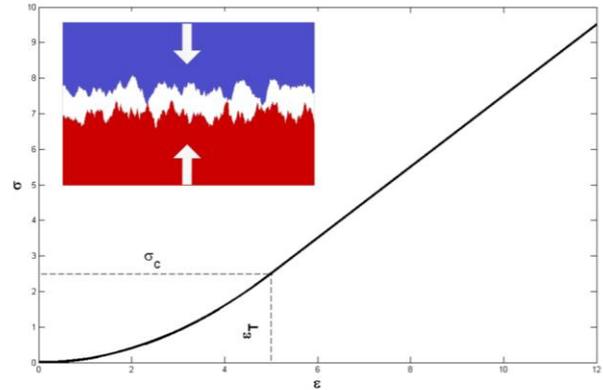

**Figure 1. Typical stress ($\sigma$)-strain($\epsilon$) curve showing transition to hard contact for elastic material occurring at a critical stress $\sigma_C$ and transition strain $\epsilon_T$.**

Under subcritical normal stresses ($\sigma < \sigma_c$), contact stiffness at the interface between two rough surfaces is governed by the geometry and distribution of asperities. Models of contact stiffness frequently employ simplified geometries to describe asperity structures. Thus, the study of contact area and forces in regular wavy surfaces exhibiting sinusoidal profiles has been the subject of several analytical interpretations (Gao et al., 2006; Johnson et al., 1985; Johnson et al., 1971) based on Hertzian contact mechanics. The earlier Greenwood and Williamson model (Greenwood and Williamson, 1966) considered uniformly spherically capped asperities exhibiting a Gaussian distribution of heights. Similar quasi-random distributions of spherical elliptical and sinusoidal asperities were considered in elastic and plastic contact mechanics models during the 1970s and 80s (Bush et al., 1976; Bush et al., 1975; McCool, 1986; Nayak, 1973). Similar surface structure simplifications are frequently employed to model the resistance of rough surfaces to shearing forces (Ciavarella et al., 1999).

The field of fractal geometry, pioneered by Mandelbrot has risen to prominence over the past several decades and finds contemporary applications in a broad range of disciplines (Mandelbrot, 1983; Mandelbrot, 1985). In particular, the tendency for naturally occurring surfaces to exhibit self-affine fractal structures has brought about an increasing focus on the significance of fractality-governed scaling behaviour of surfaces in fields of contact mechanics and tribology (Panagiotopoulos, 1992; Soare and Picu, 2007). The importance of considering surface fractality stems from the tendency of conventional roughness parameters to be dominated by longest wavelength components while higher order terms such as mean slope or curvature are dominated by shortest wavelength components.

A small range of significant experimental investigations into the significance of surface fractality on contact mechanics and tribology have been carried out (Buzio et al., 2003b; Panagouli and Iordanidou, 2013; Sun and Xu, 2005). Computational studies of the contact mechanics of fractal rough surfaces are more numerous than experimental ones and have frequently utilised geometrically self similar structures such as the type described by Weirstrass-type functions, illustrated in Figure 2(a) (Ciavarella et al., 2000; Ciavarella et al., 2004; Jackson, 2010; Warren and Krajcinovic, 1995). In such profiles identical surface structures exist at ever smaller scales, generated through iterative operations in similarity to a Koch curve or Sierpinski triangle. In contrast to self-*similar* profiles, natural surfaces are considered self-*affine,* which is to say they are similar at decreasing scales in a broader sense, exhibiting statistically similar quantities rather than being identical upon magnification (Go and Pyun, 2006). In recent years researchers have investigated more realistic simulated fractal rough surfaces, in 2D or 3D, with random self-affine roughness (Campana et al., 2011; Izquierdo et al., 2012; Li et al., 2013), such as the profile illustrated in Figure 2(b). In order to account for deterministic process governing higher level surface features, statistically self-affine surfaces with a stochastic length parameter can be simulated (Hanaor et al., 2013; Izquierdo et al., 2012; Yan and Komvopoulos, 1998). This is illustrated in Figure 2(c) and it can be seen that at the highest scale a characteristic wavelength is present such as that which could be expected from a particular mesoscopic surface process, giving an overall macroscopically flat surface.

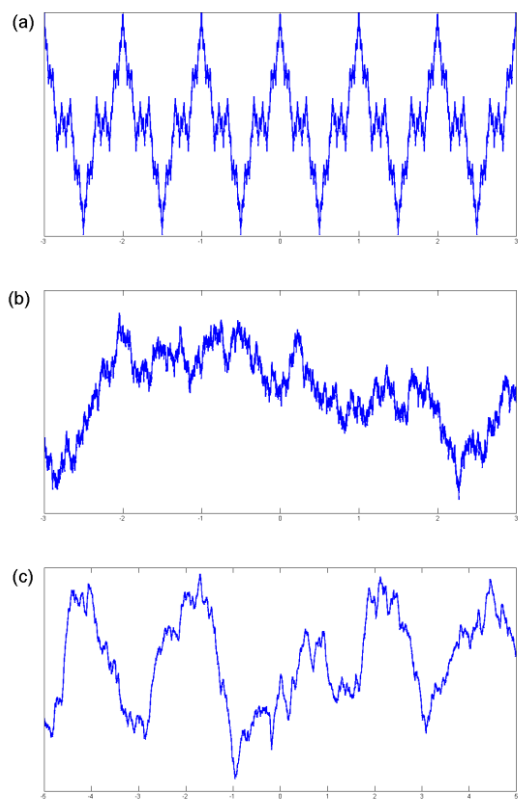

**Figure 2. Self-affine surfaces (a) Geometrically self-affine (b) Statistically self affine across all scales (c) Statistically self affine with a highest level stochastic parameter.**

Various models have been developed in order to study normal contact between randomly rough surfaces exhibiting multiscale features. Analytical and solutions for contact phenomena of theoretical multiscale self-affine surfaces have been reported using Weierstrass profiles and Archard surfaces (Ciavarella et al., 2000; Jackson, 2010) . Additionally, several continuum numerical studies have been undertaken, allowing the study of surface profiles approximating natural surfaces. Notably these have included the use of finite element methods (Hyun et al., 2004; Sahoo and Ghosh, 2007; Yastrebov et al., 2011) and Boundary Element Methods (BEM) (Pohrt and Popov, 2012; Putignano et al., 2012b). With the increase in available computational power, in order to account for the lower level interactions such as molecular scale friction, molecular dynamics based DEM models(Akarapu et al., 2011; Jerier and Molinari, 2012) have been employed. This is carried out in an attempt to account for contact phenomena at increasingly finer scales.

Below a certain stress threshold, the real contact area between rough surfaces is predicted to exhibit a linear dependence on applied load, an analytical result that is further supported by experimental data confirming this tendency (Akarapu et al., 2011; Carbone and Bottiglione, 2008, 2011; Krick et al., 2012; Putignano et al., 2012b; Xu et al., 2008). The shearing of regions of true contact, assumed to exhibit a constant shear strength, coupled with the linear dependence of contact area on applied normal load can be viewed as the origin of classical Amonton-Coulomb friction and the validation of models of normal contact mechanics of randomly rough surfaces is often carried out by testing for this linear dependence and examining the mean contact segment length (Paggi and Ciavarella, 2010). As with most studies in contact mechanics, the contact of fractal surfaces is frequently represented by truncation or half-space approximation, in which a rough surface is in contact with a rigid flat. Further, such contact events are investigated as being non-adhesive and frictionless. This simplification is adequate for problems that neglect tangential forces. However, for models to be applicable for the study of tribological interactions, it is essential to consider both normal and tangential asperity-asperity interactions across a realistic range of scales thus rendering a rough to flat approach inapplicable.

Despite the rapid development of computational capabilities, new approaches are required to evaluate the contact mechanics of forces acting at fractal surfaces. In particular this is of significance to facilitate the study of situations involving simultaneous interactions at multiple surfaces exhibiting morphing structures such as frequently encountered in the analysis of granular materials, where processes of melting, weathering and plastic deformation give rise to evolving surface fractality and consequent development of new contact mechanics and continuum scale behaviour.

In the present work we develop and investigate a computationally efficient method for the discretisation of rough surface structures and employ this tool to evaluate asperity interactions in normal and tangential orientations and thus re-examine the significance of surface fractality on contact mechanics between two simulated natural surface profiles.

## 2. Methods:

### 2.1. Generation of fractal surface profiles:

With the aim of generating representative surface profiles approximating macroscopically flat real surfaces, fractal 2-D surface profiles with a higher level stochastic parameter were generated over $1.0 \times 10^6$ (x,z) points using a variant of the previously reported method for generation of anisotropic 3-D surfaces. This method, based on Weirstrass Mandelbrot functions, is described by the following function (Ciavarella et al., 2006a; Hanaor et al., 2013; Mandelbrot, 1985; Yan and Komvopoulos, 1998) :

$$z = L^{(4-2D)} \ln \gamma \sum_{m=1}^{M} \sum_{n=0}^{n \max} \gamma^{2n(D-2)} \left\{ \cos \phi_{m,n+1} - \cos \left[ \frac{2\pi\gamma^n x}{L} \cos \left( -\pi \frac{m}{M} \right) + \phi_{m,n+1} \right] \right\}. \tag{1}$$

For 2D surface profiles (length and height) as employed here, the fractal dimension D is varied over the interval 1-2, where 1 corresponds to a smooth continuous quasi-random curve and 2 corresponds to an area filling object within set constraints (of amplitude or mean surface roughness and resolution). A stochastic length parameter, L, is included in eq. 1 to account for stochastic higher level surface features, which are present with a given maximal spacing. In real surfaces the term L represents a characteristic surface wavelength, i.e. macro-asperity spacing. Surfaces generated were scaled in the z-direction to yield a consistent mean roughness value $R_A$ across all profiles, equal to 0.5% of the profile length. The scale of the lowest level features is determined by the resolution of the simulation and this can be related to the scale of the finest roughness features in a real surface, limited by molecular scale parameters such as lattice spacing. The parameter $\gamma$ represents the density of frequencies used to construct the fractal profile. The randomised phase angle $\phi$ is given by a uniform distribution of size $M \times n_{max}$. $\phi_{m,n} = U(0, 2\pi)$. In the present work M and $n_{max}$ values of 20 were chosen as these were found to give sufficiently complex surfaces. Examples of simulated surface profiles are shown in Figure 3(a). For each contact scenario, two opposing simulated surfaces were generated with separate randomised $\phi$ sets. The equivalent 3D surfaces represented by the profiles used in the present work are illustrated in Figure 3(b). Here the surface fractal dimension ($D$) varies over the interval 2-3, where 2 corresponds to a stochastically generated smooth surface with a single level of asperities and 3 corresponds to a quasi-volume-filling object (as far as resolution allows) with features at all scales.

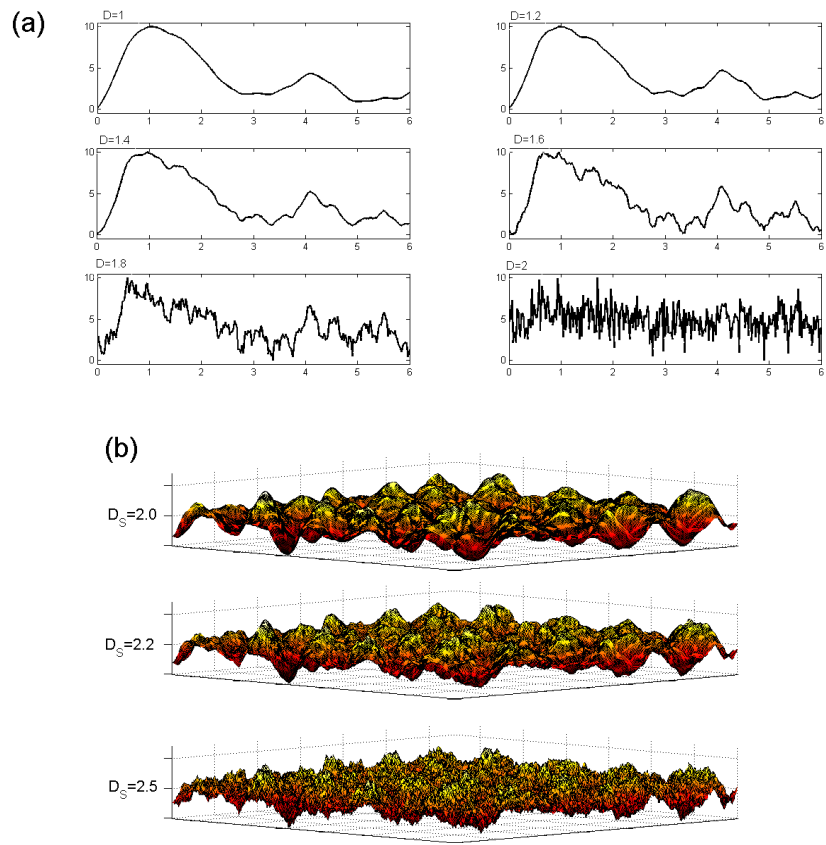

**Figure 3. (a) Representative surface profiles of varying fractal dimension D=1 to D=2 (b) Surfaces of varied fractality D=2 to D=2.5.**

### 2.2. Spline assisted discretization

Fractal surfaces are non-differentiable at all points (Sahoo and Ghosh, 2007). Hence, to discretise simulated surfaces in terms of surface normals and approximate curvature values we applied a cubic spline for smoothing generated fractal surface profiles, with 1… $i$max surface points. This approach to spline assisted discretisation yields a third order piecewise polynomial, $f(x)$ , such that the value of $F$, shown in eq. 2, is minimised by varying over the function $f(x_i)$ .

$$F = p \sum_{i=1}^{i_{max}} \left( \frac{z_i - f(x_i)}{x_0} \right)^2 + (1-p)x_0 \int (f''(x))^2 dx \tag{2}$$

For dimensional consistency here $x_0$ corresponds to unit length (e.g. 1 μm). The value of smoothness parameter $p$ is dependent on the chosen simulation resolution and therefore was set by a tightness exponent $\alpha$, using

$$p = \left( 1 + \frac{\left( dx/x_0 \right)^{\alpha}}{10} \right)^{-1} \tag{3}$$

where $dx$ is the spacing between adjacent $x$ data. Thus, high values of tightness $\alpha$ yield closer fitting interpolations, while lower values give a smoother spline. The dependence of spine fit on $\alpha$ is illustrated in Figure 2.

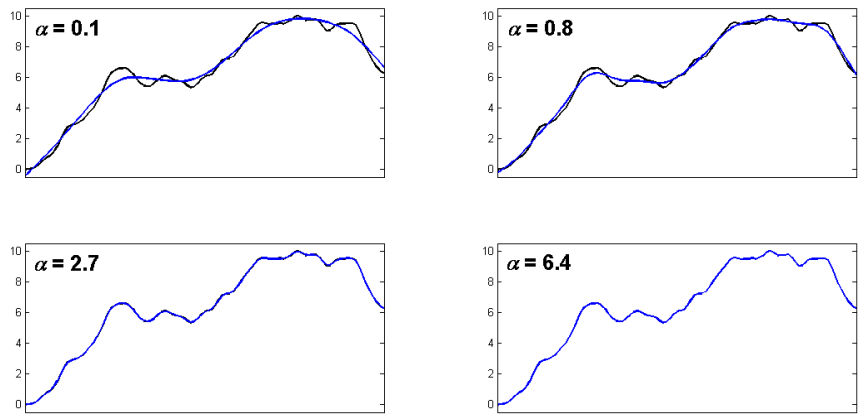

**Figure 4. Fractal surfaces, D=1.4, with spine interpolation of varied tightness exponents, $\alpha$.**

Obtaining a meaningful discretised representation of the fractal surface necessitates the use of a tightness exponent sufficiently high to represent the finest level of asperities, as determined by the profile resolution. Excessively high tightness exponent values (~$\alpha$>3) may result in erroneous interpolation of localised surface curvature yielding radii of curvature smaller than the simulation resolution. For this reason we select the a tightness exponent sufficiently high to account for surface features at all scales while not resulting in features smaller than the simulation resolution. Values in the approximate regime 1.5<$\alpha$<2.5 were found to satisfy this requirement as confirmed by consistent force displacement relationships.

### 2.3. Surface sphere interpolation:

On the basis of the spline interpolation as detailed in the previous step, we describe the upper and lower fractal surfaces in terms of local tangential 'spheres', with determined surface normals, centres and radii. These spheres are used only to convey local geometrical and material information to the contact problem, which is thus converted to a series of local contact problems between Hertzian spheres of the two surfaces.

At a given point, $(x_i, z_i)$, we utilise the coefficients of the piecewise polynomials yielded by the spline interpolation to assess local surface curvature and normals. Surface normals ($n_i$) are given by

$$n_i^l = \left\{ -f'(x_i)\left(f'(x_i)^2+1\right)^{-0.5}, \left(f'(x_i)^2+1\right)^{-0.5} \right\}^T$$
, (4.1)

$$n_i^u = \left\{ f'(x_i)\left(f'(x_i)^2+1\right)^{-0.5}, -\left(f'(x_i)^2+1\right)^{-0.5} \right\}^T$$
, (4.2)

where superscripts $l$ and $u$ respectively indicate the upper and lower surface profiles.

Radii for upper and lower tangential spheres are found as the inverse of curvature values at discrete surface points:

$$R_i^l = \frac{1}{\kappa_i^l} = \frac{\left(1+f'(x_i)^2\right)^{3/2}}{f''(x_i)}$$
(5.1)

$$R_i^u = \frac{1}{\kappa_i^u} = -\frac{\left(1+f'(x_i)^2\right)^{3/2}}{f''(x_i)}$$
(5.2)

For surface points exhibiting local concavity, $\kappa$ (and $R$) assume negative values while local convexity yields positive $\kappa$ values. Note the special case with $\kappa \to 0$ and $R \to \infty$ for a flat surface. Centres of local spheres are found by the relationships

$$\vec{O}_i^l = \vec{a}_i^l - R_i^l \vec{n}_i^l \;,$$
(6.1)

$$\vec{O}_i^u = \vec{a}_i^u - R_i^u \vec{n}_i^u \;.$$
(6.2)

Where $\vec{a}_i^l$ and $\vec{a}_i^u$ are position vectors corresponding to (x,z) coordinates at individual points of lower and upper surfaces respectively. Surface spheres and surface normals interpolation of approaching asperity of representative profiles is shown in Figure 5. The term surface sphere is used here despite the 2D simplification applied.

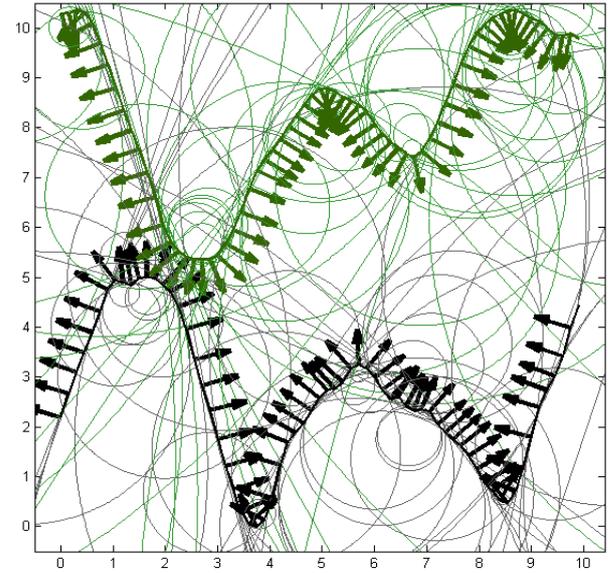

**Figure 5. Surface profiles with normals and spheres illustrated**

### 2.4. Contact model

In the evaluation of contact events the upper surface is incrementally shifted downwards, in the global normal direction. The initial contact detection criteria ($z_i^u \le z_i^l$) is applied to identify points for localised contact force analysis. For a point in contact with an opposing surface, the 'local' contact normal, $n_i^c$, between the upper and lower surfaces is a vector of unity magnitude and is given as:

$$\vec{n}_i^c = \frac{\vec{O}_i^u - \vec{O}_i^l}{\left|\vec{O}_i^u - \vec{O}_i^l\right|} = \frac{\vec{X}_i}{\left|\vec{X}_i\right|}$$
(7)

Here $\vec{X}_i$ is the vector separation of the sphere centres. This contact normal usually differs from the surface normal calculated at the grid $x=x_i$. Similarly, $\vec{t}_i^{\,c}$ is defined as the local contact tangent, a unity vector tangential to the 'local' contact.

Owing to the presence of both positive and negative curvature values and radii, contact detection is formulated by the following procedure to determine the actual centres of contact.

**Table 1. Contact selection/ rejection criteria for spline assisted discretization**

| Contact | Conditions | Acceptance | Contact position |
|---|---|---|---|
| Convex to convex | $R_i^l > 0, \;\; R_i^u > 0$ | Contact | $\vec{c}_i^{\,u} = \vec{O}_i^{\,u} - R_i^u \vec{n}_i^c$, $\vec{c}_i^{\,l} = \vec{O}_i^{\,l} + R_i^l \vec{n}_i^c$ |
| Convex to concave | $R_i^l R_i^u < 0, \;\; R_i^l + R_i^u < 0$ | Contact | $\vec{c}_i^{\,u} = \vec{O}_i^{\,u} + R_i^u \vec{n}_i^c$, $\vec{c}_i^{\,l} = \vec{O}_i^{\,l} - R_i^l \vec{n}_i^c$ |
| | $R_i^l R_i^u < 0, \;\; R_i^l + R_i^u > 0$ | No contact | - |
| Concave to concave | $R_i^l < 0, \;\; R_i^u < 0$ | No contact | - |

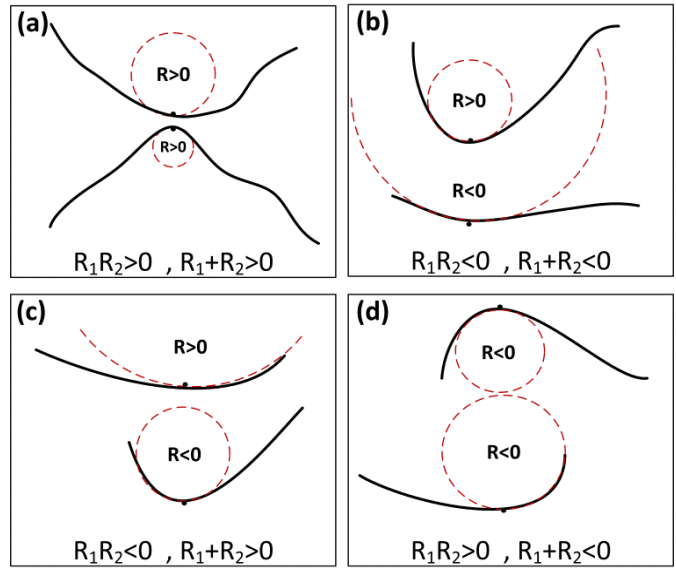

**Figure 6. Configuration of accepted contact conditions (a) and (b), and rejected contact conditions (c) and (d)**

Further, to avoid discontinuities arising from single asperities being represented by a fluctuating number of contact spheres converging at remote points, we verify the position of contact centres in the vicinity of the current meshing grid. Contact centre position is given as

$$\vec{c}_i^{\,c} = \frac{\vec{c}_i^{\,u} + \vec{c}_i^{\,l}}{2} = \frac{\vec{X}_i}{2} - \frac{R_i^u - R_i^l}{2} \vec{n}_i^c \, \mathrm{sgn}\left(R_i^u R_i^l\right).$$
(8)

With the assumption that grids are sufficiently fine to have locally smooth topology, the number of actual contacts is lower than the number of grid points. We accept contact points only when their respective centres are located inside the current domain. i.e.

$$x_i^c \in \left[ x_i - \frac{\Delta x}{2} \;,\; x_i + \frac{\Delta x}{2} \right], \qquad (9)$$

where $x_i^c$ is the global $x$-axis coordinate of the contact centre and $\Delta x$ is the grid spacing. Through this step we exclude cases where spheres' contacts occur at points distant from their respective surface profile point and thus eliminate the occurrence of contact sphere convergence at asperity peaks and consequent mesh-dependence and discontinuous force balances. In the absence of this criterion, a single peak to peak contact point may be represented by a large number of spheres corresponding to distant points, and consequently yield an erroneously high normal force contribution.

### 2.5. Force interpolation

For spline discretisation of real or simulated surfaces, accepted contact points are modelled as sections of elastically deforming Hertizan spheres. The cases for accepted contact events yield positive values for effective radius, $R^{eff}$ used by the Hertzian contact problem:

$$R_i^{eff} = \frac{2R_i^u R_i^l}{R_i^u + R_i^l} \; ;$$

$$\delta_i = R_i^u + R_i^l - \left| \vec{X}_i \right| \mathrm{sgn}(R_i^u R_i^l) \; ;$$

$$\vec{F}_i^n = \frac{\pi}{4} E^* \delta_i \vec{n}_i^c \; ; \qquad A_i = \sqrt{\delta_i R_i^{eff}} \quad (10)$$

where $\delta_i$ corresponds to sphere intrusion and $E^*$ is the effective elastic modulus. Following this approach we obtain local normal forces $\vec{F}_i^n$ and individual contact areas $A_i$ at discrete contact points present at a surface profile element of unity thickness representing a section of a three dimensional surface of fractal dimension D+1.

It should be noted that Spline assisted discretisation (SAD) can be employed to study non-Hertzian contact by utilising alternative functions of the form $F_i^n = f\left( R_i^l, R_i^u, \delta_i \right)$ incorporating plastic, visco-plastic or viscoelastic deformations. Additionally the representation of functionally graded surfaces with varying elasticity may be incorporated into SAD methods to capture the divergence between bulk and nano-scale material properties (Paggi and Zavarise, 2011).

In the present work we assume a uniform mode of (elastic) deformation at all scales. In earlier work by Majumdar and Bhusan (Bhushan and Majumdar, 1992; Majumdar and Bhushan, 1991) and Morag and Etsion (Morag and Etsion, 2007), the significance of scale dependant deformation modes were highlighted as smaller nano-scale asperities are likely to undergo plastic deformation before the larger ones. The assumptions in the present work are appropriate for conditions of low-surface penetration, such as those encountered in many-body models.

Although, in the present work, incremental surface displacement is applied only in the global normal direction ($\left| \Delta \vec{x}_i^T \right| = 0$), for each contact point we can evaluate the forces acting in the local tangential direction which in most cases will lead to a non-zero global tangential net force. However, neglecting global tangential displacement is appropriate for a laterally constrained system as frequently encountered in conditions of normal contact between macroscopically flat surfaces.

At each time step $t$ of the simulation, the tangential force at each accepted contact point $i$, modelled as two contacting spheres, is given as:

$$\vec{F}_i^t(t + \Delta t) = \vec{F}_i^t(t) + G^* \Delta \vec{x}_i^t . \qquad (11)$$

This is valid for the regime in which the magnitude of the frictional force does not exceed the magnitude of the maximum permissible shear force at a point given by the molecular scale friction $F_f$

$$F_f = \mu_0 \left| \vec{F}_i^n \right| \qquad (12)$$

Here $G^*$ is the effective shear modulus and $\Delta \vec{x}_i^t$ is the incremental displacement of point $i$ in the local tangential direction, given by the projection of the global incremental displacement on the local tangential direction, as illustrated in Figure 7. In contrast to the local normal force, the local tangential force is calculated from incremental displacements relative to the previous positions owing to the rotation and translation of asperities in the local tangential direction.

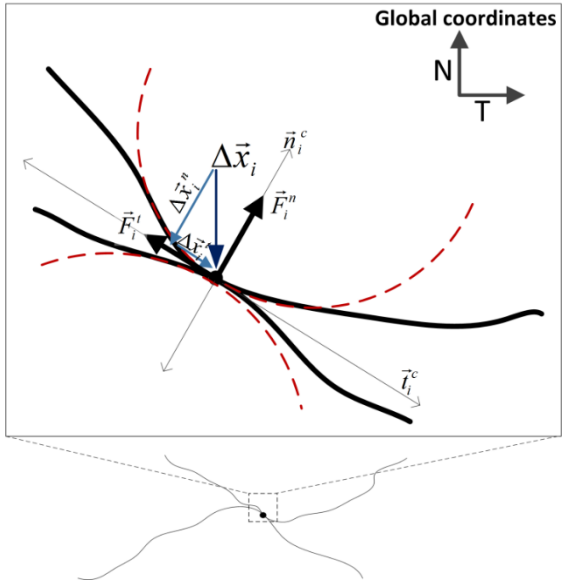

**Figure 7. Diagram of forces, normals and incremental displacements at a single contact point.**

To account for sub-asperity-scale molecular surface interactions, we apply a so-called molecular friction coefficient $\mu_0$ to account for asperity slip, which gives rise to a threshold force in the local tangential direction as

$$\text{If} \quad \left| \vec{F}_i^t \right| \geq F_f \qquad \text{then} \quad \vec{F}_{i,new}^t = \mu_0 \left| \vec{F}_i^n \right| \vec{t}_i^c \qquad (13)$$

Finally the global coordinate system force balance, comprising the total forces acting in the global normal and tangential directions ($F_N$ and $F_T$), is determined by summing the

components of individual local forces at contact points in the global coordinate system:

$$F_N = \sum \left( F_i^{n,N} + F_i^{t,N} \right), \quad F_T = \sum \left( F_i^{n,T} + F_i^{t,T} \right) \qquad (14)$$

Here $F_i^{n,N}$ and $F_i^{n,T}$ represent respectively the global normal and global tangential components of the local normal force acting at contact point $i$.

## 3. Results

The presently reported method of spline assisted discretization is applied to interpret the contact mechanics of simulated fractal surface profiles of unity thickness in normal contact with rough to rough and rough to rigid flat contact conditions considered. Figure 8 illustrates a contact event of opposing fractal surfaces while the discretised contact of individual surface features is shown in Figure 9.

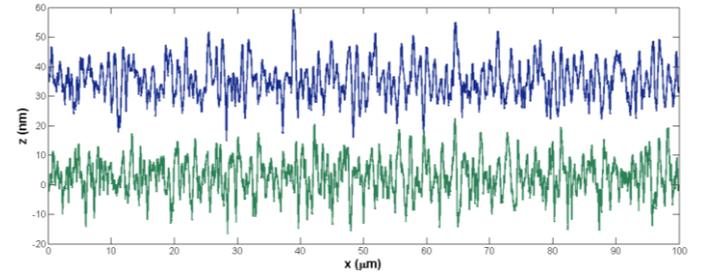

**Figure 8. Opposing fractal surface profiles at commencement of contact event**

Using a fitness exponent of $\alpha=2$ for the spline fitting described in eq. 3, fractal profiles consisting of $10^6$ data points were simulated with fractality varied over the interval D =1-1.9 following eq. 1. Relative to the stochastic length parameter L=1μm, fractal surface profiles, with contact mean roughness ($R_A$) values, were generated over a length $L_0$ totalling 100L, giving on average 100 1μm wavelengths in each profile. Surface profiles were scaled to give a controlled mean roughness ($R_A$) proportional to 1/200 relative to L, or ($R_A$=5nm). This value yields a typical aspect ratio for highest scale asperities in the simulated profiles representative of macroscopically smooth surfaces (Suh et al., 2003).

Initial conditions were controlled such that the minimal value of $z_i^u - z_i^l$ was zero at $\delta_N = 0$. The upper surface was then displaced downwards incrementally using 1000 evenly distributed increments of $\delta z$ reaching a consistent final displacement of 0.2A, where A is the amplitude of the bottom surface profile. Simulated surfaces of high fractality exhibit sharp peaks that, owing to the limited resolution of the simulation, may result in some local radii finer than the simulation resolution, consequently yielding erroneously low stiffness at those points. For this reason, a minimal surface curvature radius is applied equal to half of the profile resolution ($r_{min}=dx/2$) equating to 0.5Å, which is close proximity to a typical atomic radius, a scale below which the evaluation of Newtonian mechanical interactions is inapplicable. The use of fitness exponent $\alpha$=2 further served to minimise the occurrence of this problem.

For an equivalent modulus of 10 GPa, by evaluating individual contact events using the 2D Hertzian solution, results for the normal force acting on the profile ($F_N$ in mN ), the displacement ($\delta_N$ in nm) and the total contact area (linear projection of area, $A_S$ in units of μm) were collected and averaged over 200 independently generated surface profiles at each fractal dimension used.

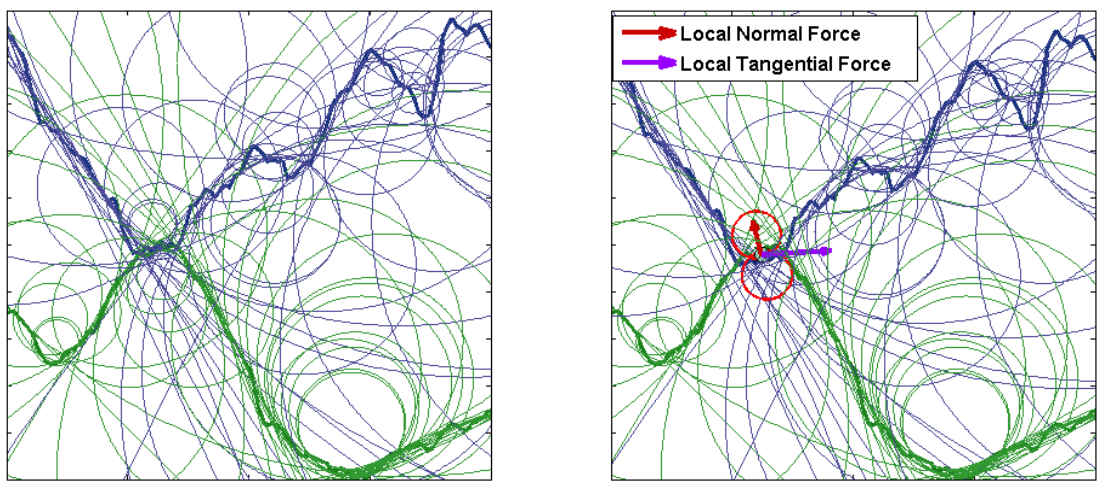

**Figure 9. SAD of contacting features of a fractal surface structure. Active contact spheres in red.**

The mean variation of normal force with displacement for each fractal dimension studied is shown in Figure 10(a). For a given surface deformation greater normal contact forces arise for surfaces with lower fractal dimensions while surfaces exhibiting greater fractality are notably more compliant. The trend of mean surface separation, corresponding to the displacement between the two mean surface planes, with increasing force is shown in Figure 10(b).Owing to the normalisation of profile heights to yield constant mean roughness values, surface pairs exhibiting larger fractal dimensions have a greater amplitude and initial mean surface separation with this disparity diminishing with increasing load.

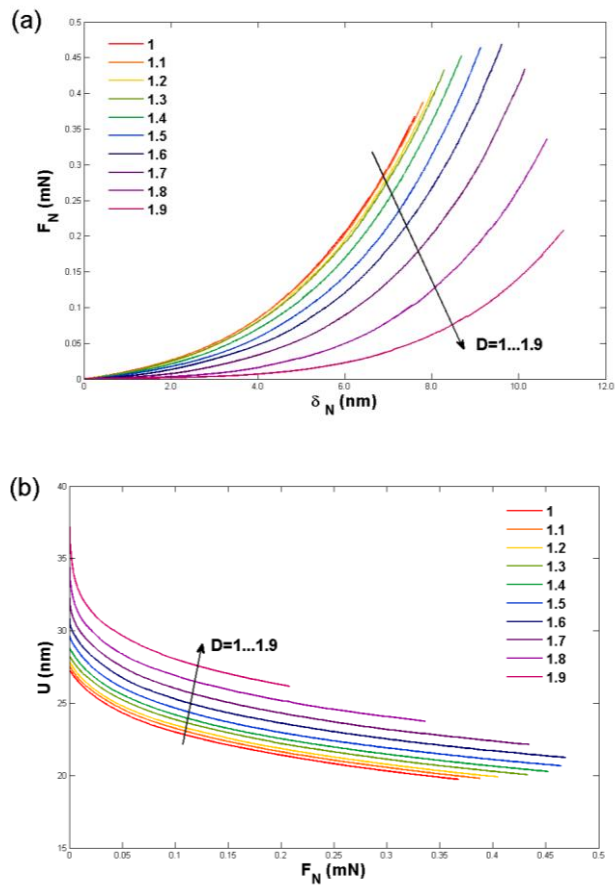

**Figure 10. (a) Overall global normal force as a function of normal displacement. (b) Mean surface separation as a function of applied normal force**

Increasing contact stiffness with vertical displacement and applied normal force is evident with all surfaces with this relationship following a power law form consistent with previous reports (Akarapu et al., 2011). The variation of non-dimensional stiffness with force, for a profile element of length/area $L_0/A_0$ as found through SAD methods, shown in Figure 11(a), is principally in agreement with reports obtained by Pohrt and Popov using a Boundary Element Methods (Pohrt and Popov, 2012) and with numerical results of Pastewka et al. (Pastewka et al., 2013) showing higher fractal dimensions exhibiting lower non-dimensional stiffness at a given load and a convergence of the stiffness/load curves for the surfaces of different fractal dimensions occuring at higher loads.

As shown in Figure 11(b) true contact area increases in a near-linear fashion with increasing applied load, complying with Greenwood-Williamson and Archard models for contact mechanics at low applied loads and exhibiting a similar trend with fractal dimension as that reported by Putignano and Ciavarella et al. (Ciavarella et al., 2006a; Ciavarella et al., 2006b; Putignano et al., 2012a). Here, for surfaces of constant mean roughness, higher fractality results in lower true contact area for a given load. Once again, this trend is theoretical at higher loads as bulk compliance means that complete contact $(A_S = A_0)$ is never reached. At low loads, deviation from linarity arises as the result of a limited number of accepted contact points in the present simulation.

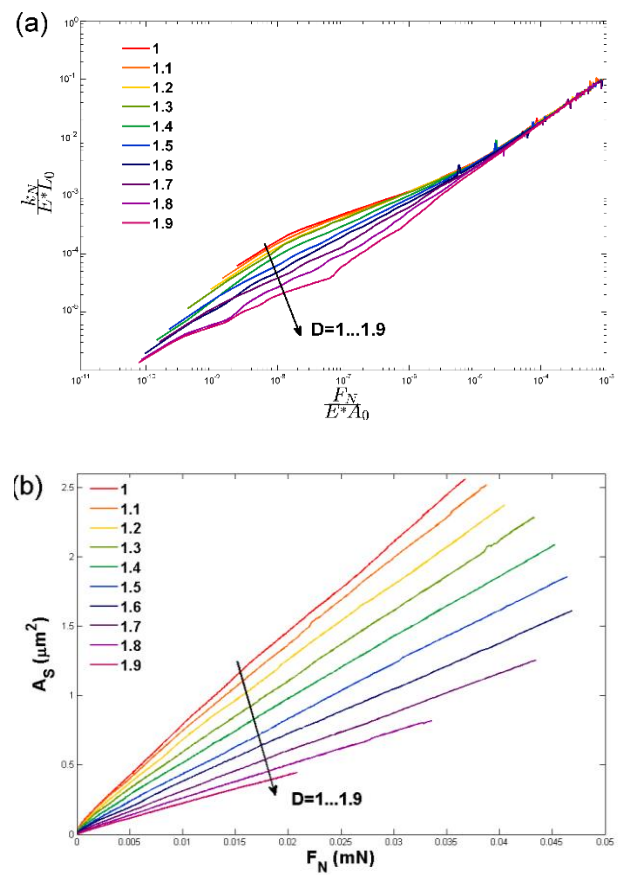

**Figure 11(a) Dimensionless normal contact stiffness vs. dimensionless normal force (b) Contact area vs applied force**

*Rough to Rough / Rough to Flat*

The present model allows the evaluation of rough to rough contact conditions in parallel to the rough to rigid flat model typically assumed. Further validation of the current SAD approach is carried out by examining the dimensionless root-mean-squared-slope normalised variation of contact area with load ($\kappa$), also known as the coefficient of proportionality (Paggi and Ciavarella, 2010). This value is calculated and compared for both rough to rough and rough to rigid-flat scenarios with differing surface fractality following:

$$\kappa = R_{Sq} A E^* F^{-1},$$ where

$$R_{Sq} = \sqrt{\left\langle \left| \nabla h \right|^2 \right\rangle} = \sqrt{\sum_{i=1}^{i=n} \left( \frac{dz_i}{dx} \right)^2 n^{-1}} \quad (15)$$

It is shown in Figure 12 that for surface profiles of fractal dimensions between 1 and 1.6 ($0.4 \leq H \leq 1$) the dimensionless values for $\kappa$ fall within the range bounded by the values predicted by Persson (Persson, 2001) and Bush (Bush et al., 1975) and are in close agreement with results obtained using Finite Element Methods by Hyun et al. (Hyun et al., 2004). In the absence of an applied global tangential load, values for the coefficient of proportionality are in agreement for both rough to rough and rough to flat scenarios. For surfaces profiles exhibiting higher fractal dimensions ($H<0.4$), not typically encountered in natural surfaces, limitations exist in the applicability of contemporary surface mechanics models. As with earlier work, the higher $\kappa$ values obtained in this regime may result from the exceedingly steep slopes ($R_{sq}>>1$) that are associated with theoretical surface profiles as the fractal dimension approaches D=2 (or surfaces with D→3).

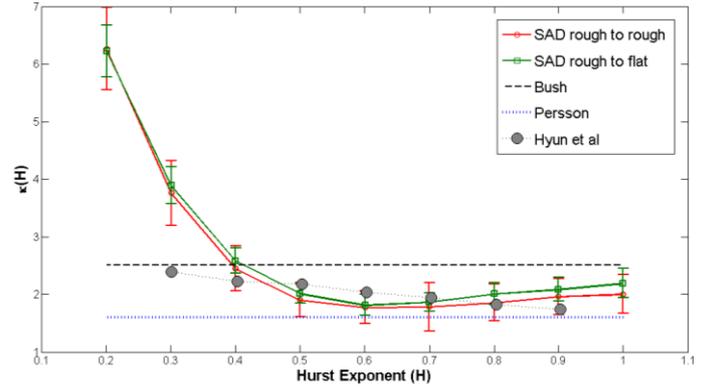

**Figure 12. Dimensionless product κ plotted vs. Hurst exponent H showing comparison with FEA models and values predicted by Bush *et al.* and Persson. (Greenwood and Williamson, 1966; Hyun et al., 2004; Persson, 2006)**

## 4. Discussion

The contact mechanics of rough surfaces exhibiting statistical self affinity across a range of scales has been the subject of much recent research using a variety of analytical methods. Relative to FEA and BEM methods the discretization approach employed in the present work has the advantage of enabling the analysis of tangential and normal interactions between two multi-scale randomly rough surfaces across in a numerical framework involving the use of reduced computational resources. The efficacy of this approach is evident through the repeated simulation (with 100 repetitions) of the normal contact of different pairs of quasi-random fractal surface profiles consisting each of $10^6$ points entailing $10^3$ displacement steps per repetition and the determination of contact area and local and global normal and tangential contact forces at each step that was achievable in a desktop computer environment in an order of ~6 hours using a dual-core CPU. Results from the present method are validated by their consistency with established relationships and experimental results with regards to the development of contact stiffness and true contact area at rough interfaces.

While the fractal surfaces studied were computationally simulated, the discretization and force interpolation methods can just as readily be applied to profilometry derived surface data (using atomic force microscopy, optical interferometry or stylus measurement). Within the framework of the presented methods a variety of asperity-contact models can be applied to account for adhesion, plasticity and viscosity. Thus the Hertzian discretization could be supplanted by Johnson-Kendall-Roberts (JKR), Maugis-Dugdale or Derjaguin-Muller-Toporov (DMT) to account for adhesive conditions or otherwise (Pietrement and Troyon, 2000) and further plasticity terms can readily be incorporated into the calculation of contact forces (Brake, 2012).

Comparable studies involving the sine-wave based discretisation of mutli-scale surface profiles were carried out by Ciavarella et al (Ciavarella et al., 2006a) building on earlier studies (Nowell and Hills, 1989), yielding similar results to the present work with respect to load/deformation behaviour and contact area development. These studies involved Weierstrass series made by the summation of up to 8 sets with distinct amplitude/wavelength pairs yielding particular contact radii. The current methods differ in their inclusion of tangential forces and the continuously scalable surface contact radii.

The topic of normal contact stiffness of fractal rough surfaces in the elastic regime was recently examined by Pohrt and Popov in a study which employed a Boundary Element type approach to study the contact of a fractal surface with a rigid flat (Pohrt and Popov, 2012; Pohrt et al., 2012). This study revealed a non-dimensionalised stiffness following a power law with relation to applied normal load and an inverse power law (negative coefficient) with relation to RMS roughness. Following logically from the positive correlation of surface fractality and RMS mean slope roughness, consistent with the results of the present work, a decreasing stiffness was observed with increasing surface fractality, with this trend diminishing at higher applied normal loads.

The majority of studies investigating normal contact stiffness, model a rough surface in contact with a rigid flat counter surface. At the micro/nano scale (below the scale of the roughness simulated) surfaces are often considered to be smooth. This approach allows the meaningful analysis of normal forces for contact problems involving pairs of surfaces exhibiting different mechanical properties or structures by utilising a hybrid surface and constitutive material properties derived from the combined structures and properties of the two surfaces of interest (Zavarise et al., 1995). Despite this, the use of a rigid flat counter surface does not allow the comprehensive consideration of asperity-asperity interactions including local normal and tangential forces and is thus of limited applicability for contact problems involving shear forces and molecular scale frictional interactions.

Persson's and Greenwood Williamson theories of contact mechanics predict a linear dependence of real contact area on applied normal load at sufficiently low loads (Greenwood and Williamson, 1966; Persson, 2006; Pohrt and Popov, 2012). This observation has been noted as the physical origin of Colulomb friction between flat surfaces and is frequently used as a benchmark to test numerical contact mechanics models. Thus both the aforementioned BEM approach, molecular dynamics (Akarapu et al., 2011) and finite element analyses (FEA) (Batrouni et al., 2002; Hyun et al., 2004; Sahoo and Ghosh, 2007) of fractal surface structures examined the linearity of the dependence of non-dimensionalised contact area on load in order to validate the numerical methods employed. Experiments involving conductance phenomena have further demonstrated the dependence of true contact area on load in fractal surfaces (Ciavarella et al., 2004). Results from the present work show that with the exception of very low loads, true contact area varies linearly with load as seen in Figure 11(b). The non linearity in the $A_C/F_N$ relationship at low loads results from the limited simulation resolution which results in only a few contact points at low loads. As shown by Greenwood and Williamson, a large number of asperities of different heights needs to be considered in order to obtain the predicted linear relation between load and contact area (Greenwood and Williamson, 1966).

The normalisation of surface area by root mean squared slope yields the value $\kappa$, also known as the coefficient of proportionality. For fractal surfaces the value of kappa is reported as a constant. Results from the present work are in reasonable agreement with finite element analysis which showed that the lower values of the Hurst dimension H, correlated to fractal dimension of 2D profiles by H=2-D, exhibit increasing $\kappa$ values. The resulting $\kappa$ values obtained in the present work by SAD are also found to largely lie between the constant values predicted by Bush *et al*. and Persson. The divergence observed at lower H values is likely the result of rapidly increasing surface slope values towards high fractality (Zavarise et al., 2007). It should be noted that such steep features are seldom encountered in natural surfaces. Owing to the absence of an applied global tangential load / displacement, in the present work $\kappa$ values are fundamentally in agreement between rough to rough and rough to flat models. However, in contrast to rigid-flat counter surface simplifications that are inadequate for the study of shear interactions at rough interfaces, the current method is of broader utility and is applicable for the analysis of frictional interactions and their correlation to surface structure.

Experimental studies in the field of normal contact mechanics of fractal surfaces confirm to a large extent the linear variation of true contact area with applied load and suggest normal surface compliance is reduced by decreasing the fractality of the contacting surfaces (Buzio et al., 2003a; Zahouani et al., 2009). Decreasing surface fractality can occur through processes of weathering, melting and abrasion and results in changes to the continuum scale mechanical behaviour exhibited by systems of multi-body contacts such as granular matter where the transition regime from soft to hard contact is of significant consequence. The currently reported method of spline assisted discretisation allows the evaluation of contact stiffness development in normal and tangential orientations in multi-body systems exhibiting evolving surface structures with potential applications for the interpretation of mechanical behaviour in granular systems.

## 5. Conclusions

A computationally efficient method for the spline assisted discretisation of multiscale surface structures has been shown to facilitate the evaluation of localised contact forces in surface-normal and surface-tangent orientations, allowing the evaluation of contact mechanics at real or simulated rough surfaces. By applying the developed method to contact events between pairs of simulated self affine rough surfaces, results showed a decreasing normal contact stiffness with increasing surface fractality and reproduced the reported linear dependence of true contact area on applied load.

By including asperity to asperity interactions, the developed methods are of broader applicability relative to simplifications of rigid flat counter surfaces and can be implemented for the analysis of contact and tribological interactions at interfaces in many-body systems exhibiting evolving surface structures.

**Acknowledgements:** Financial support for this research from the Australian Research Council through grant no. DP120104926 is gratefully appreciated.

Dorian Hanaor, Yixiang Gan, and Itai Einav. "Contact mechanics of fractal surfaces by spline assisted discretisation." *International Journal of Solids and Structures* 59 (2015): 121-131.